\documentclass[12pt,titlepage]{article}
\usepackage{amsmath}
\usepackage{amssymb}
\usepackage[dvips]{graphicx}
\usepackage{amsfonts}
\usepackage[portuguese]{babel}
\usepackage[font=small,format=plain,labelfont=bf,up,textfont=it,up]{caption}
\usepackage{appendix}

\begin{document}

\title{Estados ligados em um potencial delta duplo\\
via transformadas seno e cosseno de Fourier\thanks{To appear in Revista Brasileira de Ensino de F\'{\i}sica (Artigos Gerais)}\bigskip \\
{\small (Bound states in a double delta potential via Fourier sine and cosine transforms)}}
\author{A.S. de Castro\thanks{%
E-mail: castro@pq.cnpq.br} \\
\\
Departamento de F\'{\i}sica e Qu\'{\i}mica, \\
Universidade Estadual Paulista \textquotedblleft J\'{u}lio de Mesquita
Filho\textquotedblright, \\
Guaratinguet\'{a}, SP, Brasil}
\date{}
\maketitle

\begin{abstract}
\noindent O problema de estados ligados em um potencial delta duplo \'{e} revisto com o uso do m\'{e}todo das transformadas seno e cosseno de Fourier.
\newline
\newline 
\noindent Palavras-chave: duplo delta, estado ligado, transformadas seno e cosseno de Fourier.
\newline
\newline
\newline
\newline
\newline
\noindent {\small The problem of bound states in a double delta potential is revisited by means of Fourier sine and cosine transforms.
\newline
\newline
\noindent Keywords: double delta, bound state, Fourier sine and cosine transforms.}
\end{abstract}

\section{Introdu\c{c}\~{a}o}

Em 1927, a quantidade $\delta \left( x\right) $ definida por
\begin{equation}
\int_{-\infty }^{+\infty }dx\,\delta \left( x\right) =1,\quad \delta \left(
x\right) =0\text{\quad para\quad }x\neq 0
\end{equation}%
foi utilizada por P.A.M. Dirac para mostrar a equival\^{e}ncia entre tr\^{e}%
s diferentes formalismos da mec\^{a}nica qu\^{a}ntica \cite{dir}. Embora
seja atualmente conhecida como fun\c{c}\~{a}o delta de Dirac, tal quantidade
apareceu pela primeira vez em trabalhos de O. Heaviside j\'{a} em 1895
(veja, e.g., \cite{jac}). Conquanto $\delta \left( x\right) $ n\~{a}o seja
uma fun\c{c}\~{a}o no sentido usual da palavra, ela pode ser entendida como
o limite de uma sequ\^{e}ncia de fun\c{c}\~{o}es, e a teoria das distribui%
\c{c}\~{o}es permite que ela possa ser considerada efetivamente como uma fun%
\c{c}\~{a}o ordin\'{a}ria \cite{dis}. A utilidade da fun\c{c}\~{a}o delta
estende-se muito al\'{e}m de seu uso original em mec\^{a}nica qu\^{a}ntica e
na defini\c{c}\~{a}o de fun\c{c}\~{a}o impulsiva. \'{E} proveitosa no c\'{a}%
lculo de soma de s\'{e}ries de Fourier \cite{spi1} e de integrais \cite{spi2}%
, al\'{e}m de ser fundamental na busca de solu\c{c}\~{o}es particulares de
equa\c{c}\~{o}es diferenciais n\~{a}o-homog\^{e}neas com o m\'{e}todo da fun%
\c{c}\~{a}o de Green (veja, e.g., \cite{but}). Outrossim, a fun\c{c}\~{a}o
delta tem serventia para simular fun\c{c}\~{o}es com valores extremamente
grandes em um intervalo extremamente pequeno e para generalizar o conceito
de densidade associado com quantidades cont\'{\i}nuas para quantidades
discretas, sendo a densidade de massa de um objeto pontual um dos exemplos
mais elementares. Em muitas circunst\^{a}ncias, a raz\~{a}o da substitui\c{c}%
\~{a}o de fun\c{c}\~{o}es ordin\'{a}rias por fun\c{c}\~{o}es delta \'{e}
porque, via de regra, tais modelos tornam-se mais simples, e \`{a}s vezes
exatamente sol\'{u}veis \cite{dem}-\cite{alb1}. De fato, as fun\c{c}\~{o}es
delta t\^em sido utilizadas em uma pletora de trabalhos em situa\c{c}\~{o}es
variadas em f\'{\i}sica estat\'{\i}stica \cite{sie}, eletromagnetismo \cite%
{nam}, \'{o}tica \cite{mar}, modelos nucleares \cite{bre}-\cite{wei}, f\'{\i}%
sica do estado s\'{o}lido \cite{alb1}, \cite{alb2}, f\'{\i}sica de part\'{\i}%
culas elementares \cite{tho}, teoria do espalhamento \cite{sen}, simula\c{c}%
\~{a}o do comportamento de \'{a}tomos e mol\'{e}culas \cite{lap}-\cite{fro},
fotoioniza\c{c}\~{a}o \cite{gel}, efeito Hall quantizado \cite{pra}, f\'{\i}%
sica multidimensional \cite{bek}, efeito Casimir \cite{car}, regulariza\c{c}%
\~{a}o e renormaliza\c{c}\~{a}o em teoria qu\^{a}ntica de campos \cite{mit}.

A equa\c{c}\~{a}o de Schr\"{o}dinger com um potencial constitu\'{\i}do de
uma soma de duas fun\c{c}\~{o}es delta de Dirac, doravante denominado
potencial delta duplo, tem sido usada na descri\c{c}\~{a}o da transfer\^{e}%
ncia de um n\'{u}cleon de val\^{e}ncia durante uma colis\~{a}o nuclear \cite%
{bre} tanto quanto para para modelar as for\c{c}as de troca entre os dois n%
\'{u}cleos no \'{\i}on de hidrog\^{e}nio molecular \cite{fro}. A bem da
verdade, os estados estacion\'{a}rios de uma part\'{\i}cula em um potencial
delta duplo ocupa as p\'{a}ginas de muitos livros-texto \cite{gas}-\cite{gri}%
. Costumeiramente, os poss\'{\i}veis estados ligados s\~{a}o encontrados
pela localiza\c{c}\~{a}o dos polos complexos da amplitude de espalhamento ou
por meio de uma solu\c{c}\~{a}o direta da equa\c{c}\~{a}o de Schr\"{o}dinger
baseada na descontinuidade da derivada primeira da autofun\c{c}\~{a}o, mais
a continuidade da autofun\c{c}\~{a}o e seu bom comportamento assint\'{o}%
tico. Os estados ligados da equa\c{c}\~{a}o de Schr\"{o}dinger com
potenciais delta tamb\'{e}m tem sido alvo de investiga\c{c}\~{a}o tanto com
a transformada de Laplace \cite{lap2} quanto com a transformada exponencial
de Fourier \cite{fou2}, e no caso de potenciais constitu\'{\i}dos de fun\c{c}%
\~{o}es delta de Dirac a equa\c{c}\~{a}o de Schr\"{o}dinger independente do
tempo transmuta-se numa equa\c{c}\~{a}o alg\'{e}brica de primeira ordem para
a transformada da autofun\c{c}\~{a}o. O uso de transformadas integrais para
resolver uma equa\c{c}\~{a}o diferencial \'{e} louv\'{a}vel se a equa\c{c}%
\~{a}o transformada puder ser resolvida com maior simplicidade. No entanto,
deve-se executar a invers\~{a}o da transformada para obter a fun\c{c}\~{a}o
original do problema. Esta \'{u}ltima tarefa pode ser penosa mas \'{e}
extremamente facilitada pelo uso de tabelas de integrais.

Neste trabalho usamos as transformadas seno e cosseno de Fourier na busca de
solu\c{c}\~{o}es de estados ligados da equa\c{c}\~{a}o de Schr\"{o}dinger
com o o potencial delta duplo. A abordagem da equa\c{c}\~{a}o de Schr\"{o}%
dinger com o potencial delta duplo via transformadas seno e cosseno de
Fourier fornece uma aplica\c{c}\~{a}o adicional do m\'{e}todo de
transformadas integrais em um problema f\'{\i}sico simples que pode ser do
interesse de professores e estudantes de f\'{\i}sica matem\'{a}tica e mec%
\^{a}nica qu\^{a}ntica dos cursos de gradua\c{c}\~{a}o em f\'{\i}sica. O
procedimento adotado envolve contato com equa\c{c}\~{o}es diferenciais e o
comportamento assint\'{o}tico de suas solu\c{c}\~{o}es, fun\c{c}\~{a}o delta
de Dirac, condi\c{c}\~{o}es de contorno, paridade e extens\~{o}es sim\'{e}%
trica e antissim\'{e}trica de autofun\c{c}\~{o}es, degeneresc\^{e}ncia em
sistemas f\'{\i}sicos unidimensionais, \textit{et cetera}.

\section{As transformadas seno e cosseno de Fourier}

As transformadas seno e cosseno de Fourier de $\phi \left( x\right) $,
denotadas por $\Phi _{S}\left( k\right) $ e $\Phi _{C}\left( k\right) $
respectivamente, s\~{a}o definidas como \cite{but}-\cite{gr}
\begin{eqnarray}
\Phi _{S}\left( k\right)  &=&\mathcal{F}_{S}\left\{ \phi \left( x\right)
\right\} =\sqrt{\frac{2}{\pi }}\int_{0}^{\infty }dx\,\phi \left( x\right)
\mathrm{sen\,}kx,  \notag \\
&& \\
\Phi _{C}\left( k\right)  &=&\mathcal{F}_{C}\left\{ \phi \left( x\right)
\right\} =\sqrt{\frac{2}{\pi }}\int_{0}^{\infty }dx\,\phi \left( x\right)
\cos kx,  \notag
\end{eqnarray}%
e as transformadas inversas s\~{a}o dadas por%
\begin{equation}
\phi \left( x\right) =\left\{
\begin{array}{c}
\mathcal{F}_{S}^{-1}\left\{ \Phi \left( k\right) \right\} =\sqrt{\frac{2}{%
\pi }}\int_{0}^{\infty }dk\,\Phi _{S}\left( k\right) \mathrm{sen\,}kx, \\
\\
\mathcal{F}_{C}^{-1}\left\{ \Phi \left( k\right) \right\} =\sqrt{\frac{2}{%
\pi }}\int_{0}^{\infty }dk\,\Phi _{C}\left( k\right) \mathrm{\cos \,}kx.%
\end{array}%
\right.   \label{3}
\end{equation}%
A exist\^{e}ncia das transformadas seno e cosseno de Fourier, e de suas
inversas, {\'{e} assegurada se as fun\c{c}\~{o}es envolvidas nos integrandos
}forem absolutamente integr\'{a}veis no intervalo $[0,\infty )$, e para isto
acontecer $\phi \left( x\right) \rightarrow 0$ quando $x$ $\rightarrow
\infty $, e tamb\'{e}m $\Phi _{S}\left( k\right) $ e $\Phi _{C}\left(
k\right) $ devem tender a zero quando $k$ $\rightarrow \infty $. Neste caso,
o teorema de Parseval garante que%
\begin{equation}
\int_{0}^{\infty }dx\,|\phi \left( x\right) |^{2}=\int_{0}^{\infty
}dk\,|\Phi _{S}\left( k\right) |^{2}=\int_{0}^{\infty }dk\,|\Phi _{C}\left(
k\right) |^{2}.  \label{par}
\end{equation}%
Al\'{e}m disto, se $d\phi \left( x\right) /dx$ for tamb\'{e}m absolutamente
integr\'{a}vel valer\~{a}o as seguintes propriedades diferenciais%
\begin{eqnarray}
\mathcal{F}_{S}\left\{ \frac{d^{2}\phi \left( x\right) }{dx^{2}}\right\}
&=&-k^{2}\mathcal{F}_{S}\left\{ \phi \left( x\right) \right\} +\sqrt{\frac{2%
}{\pi }}k\phi \left( 0\right) ,  \notag \\
&& \\
\mathcal{F}_{C}\left\{ \frac{d^{2}\phi \left( x\right) }{dx^{2}}\right\}
&=&-k^{2}\mathcal{F}_{C}\left\{ \phi \left( x\right) \right\} -\sqrt{\frac{2%
}{\pi }}\phi ^{\prime }\left( 0\right) ,  \notag
\end{eqnarray}%
onde
\begin{equation}
\phi \left( 0\right) =\lim_{x\rightarrow 0_{+}}\phi \left( x\right) ,\quad
\phi ^{\prime }\left( 0\right) =\lim_{x\rightarrow 0_{+}}\frac{d\phi \left(
x\right) }{dx}.  \label{origem}
\end{equation}%
Somente quatro quatro integrais relacionadas com as transformadas seno e
cosseno de Fourier de $\mathrm{sen\,}kc$/$\left( k^{2}\pm d^{2}\right) $ e $%
\mathrm{\cos \,}kc$/$\left( k^{2}\pm d^{2}\right) $, ser\~{a}o necess\'{a}%
rios na an\'{a}lise dos estados ligados de um potencial delta duplo. Apesar
de que apenas duas integrais sejam utilizados na express\~{a}o final das
autofun\c{c}\~{o}es.

\section{Os estados ligados de um potencial delta duplo}

O potencial delta duplo sim\'{e}trico pode ser visto como o caso limite de
um potencial de po\c{c}o (ou barreira) quadrado duplo como est\'{a}
ilustrado na Figura 1, onde $L$ e $\theta $ s\~{a}o quantidades positivas.
\begin{figure}[th]
\begin{center}
\includegraphics[width=9cm, angle=0]{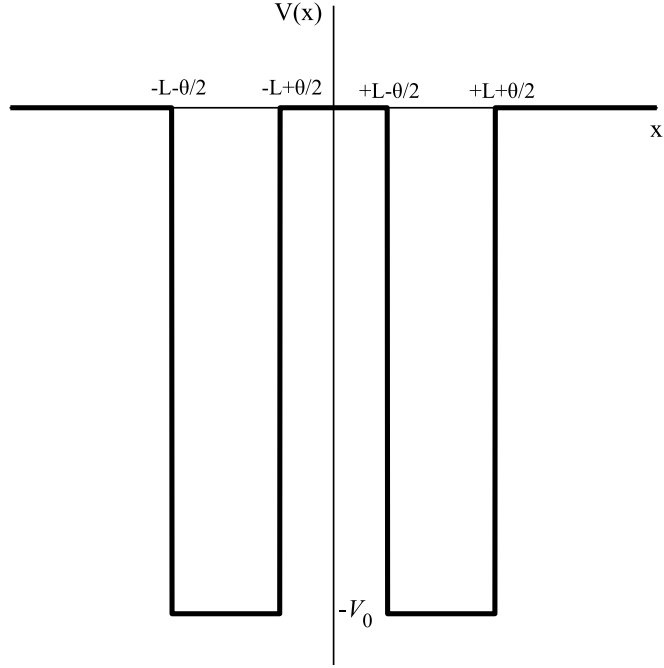}
\end{center}
\par
\vspace*{-0.1cm}
\caption{Esbo\c{c}o do po\c{c}o duplo sim\'{e}trico. Cada po\c{c}o tem
lagura $\protect\theta $ e profundidade $V_{0}$.}
\end{figure}
O limite apropriado deve ser realizado impondo $\theta \rightarrow 0$ e $%
|V_{0}|\rightarrow \infty $ de tal forma que o produto $\theta V_{0}$ permane%
\c{c}a constante. Usando as defini\c{c}\~{o}es%
\begin{equation}
\alpha =\theta V_{0}
\end{equation}
e
\begin{equation}
a=\frac{\hbar ^{2}}{2m\alpha L},\quad \kappa =\sqrt{\frac{2mE}{\hbar ^{2}}},
\label{ab}
\end{equation}%
a equa\c{c}\~{a}o de Schr\"{o}dinger independente do tempo para uma part%
\'{\i}cula de massa $m$ sujeita a um potencial delta duplo sim\'{e}trico%
\begin{equation}
V\left( x\right) =-\alpha \left[ \delta \left( x+L\right) +\delta \left(
x-L\right) \right]   \label{pot}
\end{equation}%
pode ser escrita na forma%
\begin{equation}
\frac{d^{2}\phi \left( x\right) }{dx^{2}}+\frac{1}{aL}\left[ \delta \left(
x+L\right) +\delta \left( x-L\right) \right] \phi \left( x\right) +\kappa
^{2}\phi \left( x\right) =0.  \label{eq2}
\end{equation}%
Para estados ligados, devemos procurar autofun\c{c}\~{o}es que se anulam
\`{a} medida que $|x|\rightarrow \infty $. Tendo em vista que (\ref{pot})
\'{e} invariante sob reflex\~{a}o atrav\'{e}s da origem ($x\rightarrow -x$),
se $\phi (x)$ satisfaz \`{a} equa\c{c}\~{a}o de Schr\"{o}dinger para um dado
$E$, assim acontece com $\phi (-x)$, e portanto tamb\'{e}m satisfazem as
combina\c{c}\~{o}es lineares $\phi (x)\pm $ $\phi (-x)$. Da\'{\i},
assegura-se que autofun\c{c}\~{o}es com paridades bem definidas podem ser
constru\'{\i}das, sendo suficiente concentrar a aten\c{c}\~{a}o sobre o
semieixo e impor condi\c{c}\~{o}es de contorno adicionais sobre $\phi $ na
origem. Autofun\c{c}\~{o}es em todo o eixo podem ser constru\'{\i}das
tomando combina\c{c}\~{o}es lineares sim\'{e}tricas e antissim\'{e}tricas de
$\phi $ definida no lado positivo do eixo $x$:%
\begin{equation}
\phi ^{\left( \pm \right) }\left( x\right) =[\theta \left( x\right) \pm
\,\theta \left( -x\right) ]\phi \left( |x|\right) ,  \label{todoeixo}
\end{equation}%
onde $\theta \left( x\right) $ \'{e} a fun\c{c}\~{a}o degrau de Heaviside ($1
$ para $x>0$, e $0$ para $x<0$). Estas duas autofun\c{c}\~{o}es linearmente
independentes possuem a mesma energia, ent\~{a}o, em princ\'{\i}pio, existe
uma dupla degeneresc\^{e}ncia. Por causa da continuidade da autofun\c{c}\~{a}%
o e sua derivada em $x=0$, as condi\c{c}\~{o}es de contorno sobre $\phi $ na
origem podem ser combinadas de duas formas distintas: a fun\c{c}\~{a}o par
obedece \`{a} condi\c{c}\~{a}o de Neumann homog\^{e}nea $\phi ^{\prime
}\left( 0\right) =0$, enquanto a fun\c{c}\~{a}o \'{\i}mpar obedece \`{a}
condi\c{c}\~{a}o de Dirichlet homog\^{e}nea $\phi \left( 0\right) =0$. Haja
vista que $\phi \left( x\right) $ e $d\phi \left( x\right) /dx$ s\~{a}o
nulas no infinito, temos a garantia da exist\^{e}ncia de $\mathcal{F}%
_{S}\left\{ d^{2}\phi \left( x\right) dx^{2}\right\} $ e $\mathcal{F}%
_{C}\left\{ d^{2}\phi \left( x\right) dx^{2}\right\} $ caso existam as
transformadas de $\phi \left( x\right) $ e $d\phi \left( x\right) /dx$. As
transformadas seno e cosseno de Fourier de (\ref{eq2}) resultam em equa\c{c}%
\~{o}es alg\'{e}bricas de primeiro grau:%
\begin{eqnarray}
\left( k^{2}-\kappa ^{2}\right) \Phi _{S}\left( k\right)  &=&\sqrt{\frac{2}{%
\pi }}\left[ \frac{\phi \left( L\right) }{aL}\mathrm{sen\,}kL+k\phi \left(
0\right) \right] ,  \notag \\
&& \\
\left( k^{2}-\kappa ^{2}\right) \Phi _{C}\left( k\right)  &=&\sqrt{\frac{2}{%
\pi }}\left[ \frac{\phi \left( L\right) }{aL}\cos kL-\phi ^{\prime }\left(
0\right) \right] .  \notag
\end{eqnarray}%
A condi\c{c}\~{a}o de contorno na origem dita qual o tipo de transformada
\'{e} mais apropriado: transformada seno de Fourier para uma autofun\c{c}%
\~{a}o \'{\i}mpar e transformada cosseno de Fourier para uma autofun\c{c}%
\~{a}o par. A transformada seno (cosseno) de Fourier atende \`{a} conveni%
\^{e}ncia da autofun\c{c}\~{a}o \'{\i}mpar (par) porque a transformada
inversa requer $\phi \left( 0\right) =0$ ($\phi ^{\prime }\left( 0\right) =0$%
) como pode ser depreendido de (\ref{3}). Destarte, obtemos as solu\c{c}\~{o}%
es%
\begin{eqnarray}
\Phi _{S}\left( k\right)  &=&\sqrt{\frac{2}{\pi }}\frac{\phi \left( L\right)
}{aL}\frac{\mathrm{sen\,}kL}{k^{2}-\kappa ^{2}},\quad \phi \left( -x\right)
=-\phi \left( x\right) ,  \notag \\
&&  \label{phisc} \\
\Phi _{C}\left( k\right)  &=&\sqrt{\frac{2}{\pi }}\frac{\phi \left( L\right)
}{aL}\frac{\cos kL}{k^{2}-\kappa ^{2}},\quad \phi \left( -x\right) =+\phi
\left( x\right) .  \notag
\end{eqnarray}%
Das integrais tabeladas em \cite{gr} e transcritas em (A3) e (A4) conclu%
\'{\i}mos sobre a inexist\^{e}ncia de transformadas inversas no caso em que $%
\kappa \in
\mathbb{R}
$ ($E>0$). Isto sucede porque a presen\c{c}a de $\cos \kappa x$ ou $\mathrm{%
sen\,}\kappa x$ faz com que $\phi \left( x\right) $ n\~{a}o seja
absolutamente integr\'{a}vel. Entretanto, diante das integrais tabeladas
expressas por (A1) e (A2), conclu\'{\i}mos que a exist\^{e}ncia das
transformadas inversas \'{e} assegurada para $\kappa =i\xi /L$ com $\xi >0$
de tal forma que%
\begin{equation}
E=-\frac{\hbar ^{2}\xi ^{2}}{2mL^{2}}.  \label{E}
\end{equation}%
Neste caso, a transformada inversa de (\ref{phisc}) pode ser obtida usando
os resultados tabelados expressos por (A1) e (A2). \ Assim, para $\phi
\left( -x\right) =+\phi \left( x\right) $ temos%
\begin{equation}
\phi \left( x\right) =\phi \left( 0\right) \times \left\{
\begin{array}{cc}
\cosh \frac{\xi x}{L}, & |x|\leq L, \\
&  \\
\cosh \xi \,e^{-\xi \left( |x|/L-1\right) }, & |x|\geq L,%
\end{array}%
\right.   \label{fp}
\end{equation}%
onde
\begin{equation}
\phi \left( 0\right) =\frac{\phi \left( L\right) e^{-\xi }}{a\xi }.
\label{C1}
\end{equation}%
Por outro lado, para $\phi \left( -x\right) =-\phi \left( x\right) $ temos%
\begin{equation}
\phi \left( x\right) =\frac{L}{\xi }\phi ^{\prime }\left( 0\right) \times
\left\{
\begin{array}{cc}
{\text{senh\thinspace }}\frac{\xi x}{L}, & |x|\leq L, \\
&  \\
\mathrm{\varepsilon }\left( x\right) \,{\text{senh\thinspace }}\xi \,e^{-\xi
\left( |x|/L-1\right) }, & |x|\geq L,%
\end{array}%
\right.   \label{fi}
\end{equation}%
onde $\mathrm{\varepsilon }\left( x\right) $ \'{e} a fun\c{c}\~{a}o sinal de
$x$ ($+1$ para $x>0$, e $-1$ para $x<0$), e
\begin{equation}
\phi ^{\prime }\left( 0\right) =\frac{\phi \left( L\right) e^{-\xi }}{aL}.
\label{C2}
\end{equation}%
Note que, considerando (\ref{fp}) e (\ref{fi}), podemos escrever%
\begin{equation}
\phi \left( L\right) =\left\{
\begin{array}{cc}
\phi \left( 0\right) \cosh \xi , & \phi \left( -x\right) =+\phi \left(
x\right) , \\
&  \\
\frac{L}{\xi }\phi ^{\prime }\left( 0\right) {\text{senh\thinspace }}\xi , &
\phi \left( -x\right) =-\phi \left( x\right) .%
\end{array}%
\right.   \label{c2}
\end{equation}%
A consist\^{e}ncia de (\ref{C1}), (\ref{C2}) e (\ref{c2}) requer que $\xi $
seja solu\c{c}\~{a}o das equa\c{c}\~{o}es transcendentais%
\begin{equation}
2a\xi =\left\{
\begin{array}{cc}
1+e^{-2\xi }, & \phi \left( -x\right) =+\phi \left( x\right) , \\
&  \\
1-e^{-2\xi }, & \phi \left( -x\right) =-\phi \left( x\right) .%
\end{array}%
\right.   \label{qua}
\end{equation}

A condi\c{c}\~{a}o de quantiza\c{c}\~{a}o surgiu, al\'{e}m da exig\^{e}ncia
de normalizabilidade das autofun\c{c}\~{o}es, como uma necessidade de consist%
\^{e}ncia no ajuste dos valores de $\phi \left( 0\right) $, $\phi ^{\prime
}\left( 0\right) $ e $\phi \left( L\right) $. H\'{a} de se notar que a condi%
\c{c}\~{a}o de quantiza\c{c}\~{a}o tamb\'{e}m poderia ter sido obtida pela f%
\'{o}rmula de conex\~{a}o entre $d\phi /dx$ \`{a} direita e \`{a} esquerda
de $x=L$. Tal f\'{o}rmula pode ser avaliada pela integra\c{c}\~{a}o de (\ref%
{eq2}) numa pequena regi\~{a}o em redor $x=L$, e pode ser sumarizada por%
\begin{equation}
\lim_{\varepsilon \rightarrow 0}\left. \frac{d\phi }{dx}\right\vert
_{x=L-\varepsilon }^{x=L+\varepsilon }=-\frac{\phi \left( L\right) }{aL}.
\end{equation}

\begin{figure}[th]
\begin{center}
\includegraphics[width=9cm, angle=0]{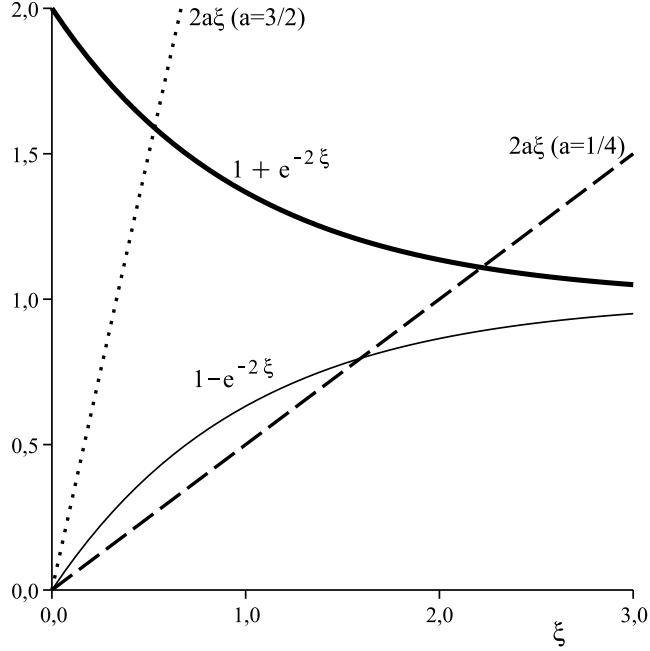}
\end{center}
\par
\vspace*{-0.1cm}
\caption{Esbo\c{c}o da condi\c{c}\~{a}o de quantiza\c{c}\~{a}o $2a\protect%
\xi =1\pm e^{-2\protect\xi }$ para $a>0$. Curva cont\'{\i}nua espessa para $%
1+e^{-2\protect\xi }$. Curva cont\'{\i}nua delgada para $1-e^{-2\protect\xi }
$. Curva pontilhada para $2a\protect\xi $ com $a=3/2$, e curva tracejada
para $2a\protect\xi $ com $a=1/4$.}
\label{Fig2}
\end{figure}

J\'{a} que as fun\c{c}\~{o}es $1+e^{-2\xi }$ e $1-e^{-2\xi }$ s\~{a}o
limitadas a valores positivos ao passo que $2a\xi $ \'{e} limitada a valores
negativos quando $a<0$, podemos inferir que n\~{a}o h\'{a} possibilidade de
solu\c{c}\~{a}o para estados ligados se $a<0$ (potencial repulsivo). Para um
potencial atrativo ($a>0$), a natureza do espectro resultante das solu\c{c}%
\~{o}es das equa\c{c}\~{o}es transcendentais (\ref{qua}) podem ser
visualizadas na Figura \ref{Fig2},
onde constam esbo\c{c}os do membro direito de (\ref{qua}), e do lado
esquerdo de (\ref{qua}) para $a=3/2$ e $a=1/4$. As curvas representadas por$%
1+e^{-2\xi }$ e $2a\xi $ sempre se interceptam. Note que $1-e^{-2\xi
}\approx 2\xi $ para $|\xi |\ll 1$ e assim a curva representada por $2a\xi $
oscula a curva representada por $1-e^{-2\xi }$ em $\xi =0$ quando $a=1$, e
estas duas curvas se interceptam em algum ponto com abscissa $\xi >0$ se e
somente se $a<1$. As abscissas das interse\c{c}\~{o}es de $1\pm e^{-2\xi }$
e $2a\xi $ correspondem aos valores permitidos de $\xi $. Pode-se depreender
da Figura 2 que sempre h\'{a} uma e somente uma solu\c{c}\~{a}o para o caso
de uma autofun\c{c}\~{a}o sim\'{e}trica, correspondendo \`{a} interse\c{c}%
\~{a}o das curvas $1+e^{-2\xi }$ e $2a\xi $. A exist\^{e}ncia de uma
adicional solu\c{c}\~{a}o, necessariamente para o caso de uma autofun\c{c}%
\~{a}o antissim\'{e}trica, sucede t\~{a}o somente quando $a<1$,
correspondendo \`{a} interse\c{c}\~{a}o das curvas $1-e^{-2\xi }$ e $2a\xi $%
. Em todo caso, os valores permitidos de $\xi $ aumentam \`{a} medida que $a$
diminui. Al\'{e}m disto, o valor permitido de $\xi $ para a solu\c{c}\~{a}o
\'{\i}mpar \'{e} menor que o valor permitido de $\xi $ para a solu\c{c}\~{a}%
o par. A rela\c{c}\~{a}o entre $E$ e $\xi $ dada por \ (\ref{E}) permite
concluir que a energia dos estados ligados diminui com o aumento de $\xi $ e
consequentemente a solu\c{c}\~{a}o com autofun\c{c}\~{a}o par corresponde ao
ub\'{\i}quo estado fundamental. Relembrando a rela\c{c}\~{a}o entre $a$, $L$
e $\alpha $ dada por (\ref{ab}), podemos concluir que a exist\^{e}ncia de
uma solu\c{c}\~{a}o \'{\i}mpar, correspondendo ao estado excitado, ocorre
somente se o potencial for suficientemente forte ($\alpha >\hbar ^{2}/2mL$).
\'{E} tamb\'{e}m instrutivo observar que, para $a<1$, o espectro torna-se
degenerado no limite $\xi \rightarrow \infty $ ($E\rightarrow -\infty $) com
$a\rightarrow 0$ ($\alpha \rightarrow \infty $).

\section{Coment\'{a}rios finais}

Este trabalho apresentou uma abordagem alternativa para a busca de estados
ligados do potencial delta duplo baseada nas transforma\c{c}\~{o}es seno e
cosseno de Fourier. Com essa abordagem a equa\c{c}\~{a}o de Schr\"{o}dinger
independente do tempo transmutou-se em equa\c{c}\~{o}es alg\'{e}bicas de
primeira ordem para as transformadas da autofun\c{c}\~{a}o. O processo de
invers\~{a}o das transformadas tornou-se amig\'{a}vel porque as integrais
envolvidas no procedimento est\~{a}o presentes em tabelas, do contr\'{a}rio
a abordagem, ainda que l\'{\i}cita, perderia o seu valor pedag\'{o}gico.

\bigskip

\bigskip

\noindent{\textbf{Agradecimentos}}

O autor \'{e} grato ao CNPq pelo apoio financeiro. Um \'{a}rbitro atencioso
contribuiu com coment\'arios e sugest\~oes pertinentes para proscrever incorre\c{c}\~{o}es na primeira vers\~ao e facetar a presente vers\~ao deste trabalho.

\appendix

\section{Integrais \'{u}teis}

\bigskip As f\'{o}rmulas 3.742.1, 3.742.3, 3.742.6 e 3.742.8 da Ref. \cite%
{gr} s\~{a}o as integrais \'{u}teis para nossos prop\'{o}sitos: {\ }%
\begin{equation}
\int_{0}^{\infty }dk\,\frac{\mathrm{sen\,}kc\,\mathrm{sen\,}kx}{k^{2}+d^{2}}=%
\frac{\pi }{2d}\times \left\{
\begin{array}{c}
e^{-cd}\,\mathrm{senh\,}dx, \\
\\
\mathrm{senh\,}cd\,\mathrm{\,}e^{-dx},%
\end{array}%
\begin{array}{c}
x<c, \\
\\
x>c,%
\end{array}%
\right.  \tag{A1}  \label{A1}
\end{equation}%
\begin{equation}
\int_{0}^{\infty }dk\,\frac{\mathrm{\cos }\,kc\,\mathrm{\cos }\,kx}{%
k^{2}+d^{2}}=\frac{\pi }{2d}\times \left\{
\begin{array}{c}
e^{-cd}\,\mathrm{\cosh }\,dx, \\
\\
\mathrm{\cosh }\,cd\,\mathrm{\,}e^{-dx},%
\end{array}%
\begin{array}{c}
x<c, \\
\\
x>c,%
\end{array}%
\right.  \tag{A2}  \label{A2}
\end{equation}%
\begin{equation}
\int_{0}^{\infty }dk\,\frac{\mathrm{sen\,}kc\,\mathrm{sen\,}kx}{k^{2}-d^{2}}=%
\frac{\pi }{2d}\times \left\{
\begin{array}{c}
\cos cd\,\,\mathrm{sen\,}dx, \\
\\
\mathrm{sen\,}cd\,\mathrm{\,}\cos dx,%
\end{array}%
\begin{array}{c}
x<c, \\
\\
x>c,%
\end{array}%
\right.  \tag{A3}  \label{A3}
\end{equation}%
\begin{equation}
\int_{0}^{\infty }dk\,\frac{\mathrm{\cos }\,kc\,\mathrm{\cos }\,kx}{%
k^{2}-d^{2}}=-\frac{\pi }{2d}\times \left\{
\begin{array}{c}
\mathrm{sen\,}cd\,\,\mathrm{\cos }\,dx, \\
\\
\mathrm{\cos }\,cd\,\mathrm{\,sen\,}dx,%
\end{array}%
\begin{array}{c}
x<c, \\
\\
x>c,%
\end{array}%
\right.  \tag{A4}  \label{A4}
\end{equation}%
com $x>0$, $c>0$ e $d\in
\mathbb{R}
$.


\newpage

\begin{thebibliography}{99}
\bibitem{dir} P.A.M. Dirac, Poc. R. Soc. London nA \textbf{113}, 621 (1927).

\bibitem{jac} J.D. Jackson, Am. J. Phys. \textbf{76}, 704 (2008).

\bibitem{dis} L. Schwartz, \textit{Theorie des distributions }(Hermann,
Paris, 1950); J.P. Marchand, \textit{Distributions }(North-Holland,
Amsterdam, 1962); I.M. Gel'fand and G.E. Shilov, \textit{Generalized
Functions }(Academic, New York, 1964).

\bibitem{spi1} M.R. Spiegel, J. Appl. Phys. \textbf{23}, 906 (1952).

\bibitem{spi2} M.R. Spiegel, J. Appl. Phys. \textbf{25}, 1302 (1954).

\bibitem{but} E. Butkov, \textit{F\'{\i}sica Matem\'{a}tica} (LTC, Rio de
Janeiro, 1988).

\bibitem{dem} Y.N. Demkov and V.N. Ostrovskii,\textit{\ Zero-Range
Potentials and Their Application in Atomic Physics} (Plenum, New York, 1988).

\bibitem{alb1} S. Albeverio \textit{et al.}, \textit{Solvable Models in
Quantum Mechanic}s (Springer-Verlag, New York, 1988).

\bibitem{sie} A.E. Siegman, Am. J. Phys. \textbf{47}, 545 (1979).

\bibitem{nam} V. Namias, Am. J. Phys. \textbf{45}, 624 (1977); D.T.
Gillespie, Am. J. Phys. \textbf{51}, 520 (1983); D.G. Hall, Am. J. Phys.
\textbf{63}, 508 (1995).

\bibitem{mar} D. Marcuse, Theory of Dielectric Optical Waveguides (Academic,
New York, 1974).

\bibitem{bre} G. Breit, \textit{Proc. Second Conf. Reactions between Complex
Nuclei, Gatlinburg}, p.1 (Wiley, New York, 1960); G. Breit, Ann. Phys.
(N.Y.) \textbf{34}, 377 (1965).

\bibitem{wei} R. de L. Kronig and W.G. Penney, Proc. Roy. Soc. London A
\textbf{130}, 499 (1931); S. Weinberg, Phys. Lett. B \textbf{251}, 288
(1990); S. Weinberg, Nucl. Phys. B \textbf{363}, 3 (1991).

\bibitem{alb2} S. Albeverio and R. Hoegh-Kronh, J. Oper. Theor. \textbf{6},
313 (1981); J.E. Avron \textit{et al.}, Phys. Rev. Lett. \textbf{72}, 896
(1994).

\bibitem{tho} E.U. Condon, Phys. Rev. \textbf{49}, 459 (1936); C. Thorn,
Phys. Rev. D \textbf{19}, 639 (1979).

\bibitem{sen} I.R. Lapidus, Am. J. Phys. \textbf{37}, 930 (1969); I.R.
Lapidus, Am. J. Phys. \textbf{37}, 1064 (1969); E.Kujawski, Am. J. Phys.
\textbf{39}, 1248 (1971); P. Senn, Am. J. Phys. \textbf{56}, 916 (1988).

\bibitem{lap} C.M. Rosenthal, J. Chem. Phys. \textbf{55}, 2474 (1971); I.R.
Lapidus, Am. J. Phys. \textbf{43}, 790 (1975).

\bibitem{fro} A.A. Frost, J. Chem. Phys. \textbf{25}, 1150 (1956).

\bibitem{gel} S. Geltman, J. Phys. B \textbf{10}, 831 (1977); E.J. Austin,
J. Phys. B \textbf{12}, 4045 (1979); H. Fearn and W.E. Lamb Jr., Phys. Rev.
A \textbf{43}, 2124 (1991).

\bibitem{pra} R.E. Prange, Phys. Rev. B \textbf{23}, 4802 (1981).

\bibitem{bek} E. Demiralp and H. Beker, J. Phys. A \textbf{36}, 7449 (2003);
S. Giardino, Rev. Bras. Ens. Fis. \textbf{35}, 3307 (2013).

\bibitem{car} L.A.S. Carrilho and J.A. Nogueira, Rev. Bras. Ens. Fis.
\textbf{31}, 2311 (2009).

\bibitem{mit} I. Mitra \textit{et al.}, Am J. Phys. 66, 1101 (1998); R.M.
Cavalcanti, Rev. Bras. Ens. Fis. \textbf{21}, 336 (1999).

\bibitem{gas} S. Gasiorowicz, \textit{F\'{\i}sica Qu\^{a}ntica} (Guanabara
Dois, Rio de Janeiro 1974).

\bibitem{coh} C. Cohen-Tannoudji \textit{et al.}, \textit{Quantum Mechanics}%
, Vol.1 (Hermann, Paris 1977).

\bibitem{gal} A. Galindo and R. Pascual, \textit{Quantum Mechanics I}
(Springer-Verlag, Berlin 1990).

\bibitem{got} K. Gottfried and T.-M. Yan, \textit{Quantum Mechanics:
Fundamentals}, 2nd. ed. (Springer, New York, 2003).

\bibitem{tam} K. Tamvakis, \textit{Problems \& Solutions in Quantum Mechanics%
} (Cambridge University Press, Cambridge, 2005).

\bibitem{rob} R.W. Robinett, \textit{Quantum Mechanics}, 2nd. ed. (Oxford
University Press, Oxford, 2006).

\bibitem{gri} D.J. Griffiths, \textit{Mec\^{a}nica Qu\^{a}ntica}, 2a. ed.
(Pearson Prentice Hall, S\~{a}o Paulo, 2011).

\bibitem{lap2} A.S. de Castro, Rev. Bras. Ens. Fis. \textbf{34}, 4301 (2012).

\bibitem{fou2} A.S. de Castro, Rev. Bras. Ens. Fis. \textbf{34}, 4304 (2012).

\bibitem{gr} I.S. Gradshteyn and I.M. Ryzhik, \textit{Table of Integrals,
Series, and Products}, A. Jeffrey and D. Zwillinger (eds.), 7th ed.
(Academic Press, New York, 2007).
\end{thebibliography}
\end{document}